\documentstyle[12pt]{article}

\def\ti#1 {\begin{center} \vspace*{2.98074cm} \baselineskip=17pt {\large #1}
\end{center}}

\def\au#1 {\begin{center} #1 \end{center}}

\def\fig#1 #2 #3 #4 {\begin{figure} \vspace{#3pt} \caption[#1]{#4} \label{#1}
\end{figure}}

\def\ben{\begin{enumerate}}
\def\een{\end{enumerate}}

\def\ess{\hskip.444444em plus .499997em minus .037036em}
\def\mss{\hskip.333333em plus .208331em minus .088889em}
\def\sen{\hbox{\scriptsize--}}
\def\eV{e\kern-.10emV }
\def\eVcm{e\kern-.10emV\kern-.15em,\mss}
\def\eVp{e\kern-.10emV\kern-.15em.\ess}
\def\eVpr{e\kern-.10emV) }

\begin{document}

\setcounter{totalnumber}{2}
\setcounter{topnumber}{2}
\setcounter{bottomnumber}{2}
\renewcommand{\topfraction}{1.0}
\renewcommand{\bottomfraction}{1.0}
\renewcommand{\textfraction}{0.0}

\vspace*{-0.387cm}
\ti{MACROSCOPIC-MICROSCOPIC MASS MODELS}

\au{\underline{J. Rayford Nix}\\
Theoretical Division, Los Alamos National Laboratory\\
Los Alamos, New Mexico 87545, USA\\
and\\
Peter M\"oller\\
Center for Mathematical Sciences, University of Aizu\\
Aizu-Wakamatsu, Fukushima 965-80, Japan}
\vspace{44.848pt}

\begin{quotation}
{\bf Abstract:}  We discuss recent developments in macroscopic-microscopic mass
models, including the 1992 finite-range droplet model, the 1992
extended-Thomas-Fermi Strutinsky-integral model, and the 1994 Thomas-Fermi
model, with particular emphasis on how well they extrapolate to new regions of
nuclei.  We also address what recent developments in macroscopic-microscopic
mass models are teaching us about such physically relevant issues as the
nuclear curvature energy, a new congruence energy arising from a
greater-than-average overlap of neutron and proton wave functions, the nuclear
incompressibility coefficient, and the Coulomb redistribution energy arising
from a central density depression.  We conclude with a brief discussion of the
recently discovered rock of metastable superheavy nuclei near $^{272}$110 that
had been correctly predicted by macroscopic-microscopic models, along with a
possible new tack for reaching an island near $^{290}$110 beyond our present
horizon.
\end{quotation}

\baselineskip=18pt
\vspace{11pt}

\noindent{\bf 1.  Introduction}

The accurate calculation of the ground-state mass and deformation of a nucleus,
specified by its proton number $Z$ and neutron number $N$, remains a
fundamental challenge of nuclear theory.  Approaches developed over the years
to achieve this difficult yet all-important goal span a broad spectrum:  (1)
fully selfconsistent microscopic theories starting with an underlying
nucleon-nucleon interaction, (2) macroscopic-microscopic models utilizing
calculated shell and pairing corrections, (3) mass formulas with empirical
shell terms whose parameters are extracted from adjustments to experimental
masses, (4) algebraic expressions based on the nuclear shell model, and (5)
neural networks.

At the most fundamental of the above levels, microscopic theories have seen
progress in both the nonrelativistic Hartree-Fock approximation$^{\ref{QF})}$
and more recently the relativistic mean-field
approximation.$^{\ref{SW}\sen\ref{LSHKR})}$\ess  Important steps have also been
taken to construct an effective Lagrangian for use in such theories based on
the chiral symmetry of quantum chromodynamics.$^{\ref{W}\sen\ref{MF})}$\ess
Microscopic theories offer great promise for the future, but so far have been
applied only to limited regions of mainly spherical nuclei, with accuracies
that are typically a few M\eVp  For example, the average absolute deviation
between the calculated and experimental masses of 32 spherical nuclei that were
considered in one relativistic treatment$^{\ref{NHM})}$ is 2.783 M\eVp

At the second level of fundamentality, the macroscopic-microscopic
method---where the smooth trends are obtained from a macroscopic model and the
local fluctuations from a microscopic model---has been used in several recent
global calculations.  We will concentrate here on three such
calculations,$^{\ref{MNMS}\sen\ref{MSc})}$\mss with particular emphasis on how
well they extrapolate to new regions of nuclei, but will also include an
example of each of the remaining approaches.$^{\ref{vGHT}\sen\ref{GCPB})}$\ess
We will then discuss some new physical insight provided by
macroscopic-microscopic mass models, and conclude with their prediction of the
recently discovered rock of metastable superheavy nuclei near $^{272}$110 and
of an island near $^{290}$110 beyond our present horizon.

\bigskip
\bigskip

\noindent{\bf 2.  Finite-Range Droplet Model}

\fig me adndt/fig1cor 253 {Comparison of experimental and calculated
microscopic corrections for the original 1654 nuclei included in the 1992
adjustment of the finite-range droplet model.$^{\ref{MNMS},\ref{MNK})}$\ess The
difference between these two quantities shown in the bottom part of the figure
is equivalent to the difference between experimental and calculated masses.}

In the finite-range droplet model, which takes its name from the macroscopic
model that is used, the microscopic shell and pairing corrections are
calculated from a realistic, diffuse-surface, folded-Yukawa single-particle
potential by use of Strutinsky's method.$^{\ref{S})}$\ess  In 1992 we made a
new adjustment of the constants of an improved version of this
model$^{\ref{MNMS},\ref{MNK})}$ to 28 fission-barrier heights and to 1654
nuclei with $N,Z \ge 8$ ranging from $^{16}$O to $^{263}$106 whose masses were
known experimentally in 1989.$^{\ref{A})}$\ess  The resulting microscopic
corrections are shown in Fig.~\ref{me}.

The improvements include minimization of the nuclear potential energy of
deformation with respect to $\epsilon_3$ and $\epsilon_6$ shape degrees of
freedom in addition to the usual $\epsilon_2$ and $\epsilon_4$ deformations,
use of the Lipkin-Nogami extension of the BCS method for calculating the
pairing correction, use of a new functional form and optimized constant for the
effective-interaction pairing gap, use of ground-state single-particle levels
calculated for each nucleus individually, use of an eighth-order Strutinsky
shell correction with basis functions containing 12 oscillator shells, and
inclusion of a zero-point energy in the quadrupole degree of freedom only.
This model has been used to calculate the ground-state masses, deformations,
odd-particle spins, pairing gaps, separation energies, and other properties for
8979 nuclei with $N,Z \ge 8$ ranging from $^{16}$O to $^{339}136$ and extending
from the proton drip line to the neutron drip
line.$^{\ref{MNMS},\ref{MNK})}$\ess

For the original 1654 nuclei included in the adjustment, the theoretical error,
determined by use of the maximum-likelihood method with no contributions from
experimental errors,$^{\ref{MNMS},\ref{MNK})}$\mss is 0.669 M\eVp  As can be
seen in the bottom part of Fig.~\ref{me}, some large systematic errors are
still present for light nuclei but decrease significantly for heavier nuclei.
The decrease in theoretical error with increasing mass number is presented
quantitatively in Fig.~\ref{te}.  Each solid circle gives the theoretical error
for a region extending 24 mass units below it and 25 mass units above it.
These points are well represented by the function $8.62 \; {\rm M\eV}/
A^{0.57}$ shown by the solid line.

\fig te adndt/fig7nn 253 {Theoretical error as a function of mass number for
the original 1654 nuclei included in the 1992 adjustment of the finite-range
droplet model.$^{\ref{MNMS},\ref{MNK})}$\ess  The dashed line shows the
constant value 0.669 M\eV appropriate to the entire region of nuclei
considered.}

\bigskip
\bigskip

\noindent{\bf 3.  Extrapolateability to New Regions of Nuclei}

Between 1989 and 1993, the masses of 217 additional nuclei heavier than
$^{16}$O have been measured,$^{\ref{AW})}$\mss which provides an ideal
opportunity to test the ability of mass models to extrapolate to new regions of
nuclei whose masses were not included in the original adjustment.
Figure~\ref{frdm} shows as a function of the number of neutrons from
$\beta$-stability the individual deviations between these newly measured masses
and those predicted by the 1992 finite-range droplet model.  The theoretical
error for these 217 newly measured masses is 0.642 M\eVcm corresponding to a
{\it decrease\/} of 4\%.  This model can therefore be extrapolated to new
regions of nuclei with considerable confidence.

\fig frdm adndt2/devmfrdm 235 {Deviations between experimental and calculated
masses for 217 new nuclei whose masses were not included in the 1992 adjustment
of the finite-range droplet model.$^{\ref{MNMS},\ref{MNK})}$\ess}

Analogous deviations are shown in Fig.~\ref{etfsi} for the 1992 Thomas-Fermi
Strutinsky-integral model (version 1) of Aboussir, Pearson, Dutta, and
Tondeur.$^{\ref{APDT})}$\ess  In this model, the macroscopic energy is
calculated for a Skyrme-like nucleon-nucleon interaction by use of an extended
Thomas-Fermi approximation.  The shell correction is calculated from
single-particle levels corresponding to this same interaction by use of a
Strutinsky-integral method, and the pairing correction is calculated for a
$\delta$-function pairing interaction by use of the conventional BCS
approximation.  The constants of the model were determined by adjustments to
the ground-state masses of 1492 nuclei with mass number $A \ge 36$, which
excludes the troublesome region from $^{16}$O to mass number $A = 35$.  The
theoretical error corresponding to 1538 nuclei whose masses were known
experimentally$^{\ref{A})}$ at the time of the original adjustment is 0.726
M\eVp  The theoretical error for 210 newly measured masses$^{\ref{AW})}$ for
nuclei with $A \ge 36$ is 0.810 M\eVcm corresponding to an increase of 12\%.
Some caution should therefore be exercised when extrapolating this model to new
regions of nuclei.

\fig etfsi adndt2/devmetf 253 {Deviations between experimental and calculated
masses for 210 new nuclei whose masses were not included in the 1992 adjustment
of the extended-Thomas-Fermi Strutinsky-integral
model.$^{\ref{APDT})}$\medskip}

\fig tf arles/tf94 253 {Deviations between experimental and calculated masses
for 217 new nuclei whose masses were not included in the 1994 adjustment of the
Thomas-Fermi model.$^{\ref{MS}\sen\ref{MSc})}$\ess}

Similar results are shown in Fig.~\ref{tf} for the 1994 Thomas-Fermi model of
Myers and Swiatecki.  In this model,$^{\ref{MS}\sen\ref{MSc})}$\mss the
macroscopic energy is calculated for a generalized Seyler-Blanchard
nucleon-nucleon interaction by use of the original Thomas-Fermi approximation.
For $N,Z \ge 30$ the shell and pairing corrections were taken from the 1992
finite-range droplet model, and for $N,Z \le 29$ a semi-empirical expression
was used.  The constants of the model were determined by adjustments to the
ground-state masses of the same 1654 nuclei with $N,Z \ge 8$ ranging from
$^{16}$O to $^{263}$106 whose masses were known experimentally in 1989 that
were used in the 1992 finite-range droplet model.  The theoretical error
corresponding to these 1654 nuclei is 0.640~M\eVp  The reduced theoretical
error relative to that in the 1992 finite-range droplet model arises primarily
from the use of semi-empirical microscopic corrections in the extended
troublesome region \linebreak $N,Z \le 29$ rather than microscopic corrections
calculated more fundamentally.  The theoretical error for 217 newly measured
masses$^{\ref{AW})}$ is 0.737 M\eVcm corresponding to an increase of 15\%.
Some caution should again be exercised when extrapolating this model to new
regions of nuclei.

These theoretical errors are summarized in Table~\ref{extrap}, where we also
include---because of its frequent use in astrophysical calculations---the 1976
mass formula of von~Groote, Hilf, and Takahashi.$^{\ref{vGHT})}$\ess  For this
mass formula with empirical shell terms whose parameters are extracted from
adjustments to experimental masses, the theoretical error for 217 newly
measured masses$^{\ref{AW})}$ increases by 104\% relative to the theoretical
error for 1323 nuclei whose masses were known experimentally in 1977.  This
formula extrapolates to new regions of nuclei very poorly and is therefore
inappropriate for use in modern-day astrophysical calculations.

For the recent mass formula of Duflo and Zuker,$^{\ref{DZ})}$\mss which is an
example of an algebraic expression based on the nuclear shell model, the
increase in root-mean-square error for newly measured masses relative to that
for masses included in the original adjustment is 32\%.  Finally, for one
particular neural network of Gernoth, Clark, Prater, and Bohr that had not been
pruned for improved extrapolateability,$^{\ref{GCPB})}$\mss the increase in
root-mean-square error for newly measured masses relative to that for masses
included in the original training is 772\%.  These last two examples are not
included in Table~\ref{extrap} because the regions of newly measured masses are
different from those in the table and because we have available only the
root-mean-square errors, which are contaminated by contributions from
experimental errors.

\begin{table}
\caption[extrap]{Extrapolateability to New Regions of Nuclei.}
\label{extrap}
\vspace{-2.233pt}
\begin{center}
{\begin{tabular}{lcccccccc}
\hline\hline \vspace{-10.150pt} \\
& & \multicolumn{2}{c}{Original nuclei} & & \multicolumn{2}{c}{New nuclei} & &
\\[-0.200pt]
\cline{3-4}\cline{6-7}\\[-9.750pt]

Model & & ${N}_{\rm nuc}$ & Error & & ${N}_{\rm nuc}$ & Error & & Error \\

& & & (M\eVpr & & & (M\eVpr & & \hspace{0pt} ratio \vspace{1.275pt} \\
\hline \vspace{-9.675pt} \\

FRDM (1992) & & 1654 & 0.669 & & 217 & 0.642 & & 0.96 \\[6.5pt]

ETFSI-1 (1992) & & 1538 & 0.726 & & 210 & 0.810 & & 1.12 \\[6.5pt]

TF (1994) & & 1654 & 0.640 & & 217 & 0.737 & & 1.15 \\[6.5pt]

v.\ Groote (1976) & & 1323 & 0.629 & & 217 & 1.284 & & \hspace{0pt} 2.04
\vspace{1.275pt} \\
\hline\hline
\end{tabular}}
\vspace{-11pt}
\end{center}
\end{table}

\bigskip
\bigskip

\noindent{\bf 4.  New Physical Insight}

Recent developments in macroscopic-microscopic mass models are also providing
new insight in several areas, including the nuclear curvature energy, a new
congruence energy, the nuclear incompressibility coefficient, and the Coulomb
redistribution energy.  Myers and Swiatecki offer a simple resolution of the
long-standing nuclear-curvature-energy anomaly$^{\ref{SBNS})}$ in terms of
their 1994 Thomas-Fermi model.$^{\ref{MS}\sen\ref{MSc})}$\ess  This model is
characterized by a curvature-energy constant $a_3$ = 12.1 M\eV but nevertheless
adequately reproduces nuclear ground-state masses through the counteraction of
terms that are of still higher order in $A^{-1/3}$.  The fission barriers of
medium-mass nuclei calculated with such a large curvature-energy constant have
in the past been significantly higher than experimental values, but in their
view the shape dependence of a new congruence energy arising from a
greater-than-average overlap of neutron and proton wave
functions$^{\ref{MSd})}$ resolves this difficulty.

The value of the incompressibility coefficient $K$ adopted in the 1992
finite-range droplet model$^{\ref{MNMS},\ref{MNK})}$ from a variety of
considerations is 240 M\eVp  An adjustment of $K$ in this model to optimally
reproduce both ground-state masses and fission-barrier heights gives 243 M\eVcm
although this adjustment is not able to rule out values of $K$ in the range
from somewhat below 200 M\eV to about 500 M\eVp  The value of $K$ determined in
the 1992 extended-Thomas-Fermi Strutinsky-integral model$^{\ref{APDT})}$ is
234.7 M\eV and that determined in the 1994 Thomas-Fermi
model$^{\ref{MS}\sen\ref{MSc})}$ is 234 M\eVp  However, these values are also
subject to large uncertainties.

The inclusion of $\epsilon_3$ and $\epsilon_6$ shape degrees of freedom in the
finite-range droplet model permitted the isolation of the Coulomb
redistribution energy, in which the nuclear ground-state mass is lowered
through the development of a central depression in the nuclear charge
density.$^{\ref{MNMSb})}$\ess  This effect also appears naturally in other
macroscopic models such as the 1992 extended-Thomas-Fermi model$^{\ref{APDT})}$
and the 1994 Thomas-Fermi model$^{\ref{MS}\sen\ref{MSc})}$.\ess  The magnitude
of this energy, which is several M\eV for a heavy nucleus, increases strongly
with increasing proton number.  Consequently, models that do not take it into
account are likely to be seriously in error when making predictions for
superheavy nuclei.

\bigskip
\bigskip

\noindent{\bf 5.  Rock of Metastable Superheavy Nuclei}

\fig ld arles/110272 313 {Dependence of single-proton and single-neutron
energies upon deformation$^{\ref{MNMS},\ref{MNK})}$ for the metastable
superheavy nucleus $^{272}$110.}

Several macroscopic-microscopic mass models$^{\ref{MLN}\sen\ref{CHN})}$ predict
a rock of deformed metastable superheavy nuclei near $^{272}$110, in addition
to an island of superheavy nuclei associated with the more familiar spherical
magic proton number $Z$ = 114 and neutron number $N$ = 184.  As illustrated in
Fig.~\ref{ld}, the underlying physical origin of this rock of deformed
superheavy nuclei is the development of large gaps in the single-particle
energies at proton number $Z$ = 104, 106, 108, and 110 and at neutron number
$N$ = 162 and 164 for prolate deformations in the vicinity of $\epsilon_2$ =
0.2.

Because superheavy nuclei can decay by either spontaneous fission, alpha decay,
or beta decay (including electron capture), all of these decay modes must be
considered when calculating their half-lives.  The spontaneous-fission
half-life depends upon both the fission barrier and the inertia for motion in
the fission direction.  This inertia can be calculated by use of either a
microscopic cranking model or a semi-empirical relationship involving the
irrotational inertia.  The half-lives for alpha decay and beta decay depend
primarily upon the corresponding energy releases $Q_\alpha$ and $Q_\beta$,
which are given by appropriate differences of ground-state masses.  When all
decay modes are considered, the longest-lived spherical superheavy nuclei are
those near $^{290}$110, with predicted half-lives of the order of years.  The
deformed superheavy nuclei near $^{272}$110 are predicted to have shorter
half-lives of the order of milliseconds.$^{\ref{MLN}\sen\ref{CHN})}$

Deformed metastable superheavy nuclei near $^{272}$110 have recently been
discovered$^{\ref{H+}\sen\ref{Ob})}$ through reactions of the type illustrated
in Fig.~\ref{me3d}.  In the cold-fusion reactions involving nearly magic
spherical targets utilized at the Gesellschaft f\"ur Schwerionenforschung in
Darmstadt, Germany, the excitation energy is sufficiently low that the compound
nuclei produced can de-excite to their ground states by the emission of a
single neutron.$^{\ref{H+}\sen\ref{H})}$\ess  In contrast, in the hot-fusion
reactions involving deformed targets utilized at the Joint Institute for
Nuclear Research in Dubna, Russia by a Dubna-Livermore team, the excitation
energy is sufficiently high that the compound nuclei produced must de-excite to
their ground states by the emission of some five
neutrons.$^{\ref{L+}\sen\ref{Ob})}$\ess

Ingenius new tacks will be required to reach the island of spherical superheavy
nuclei near $^{290}$110 that is predicted to lie beyond our present horizon.
One possibility involves the use of prolately deformed targets and projectiles
that also possess large negative hexadecapole moments, which leads to large
waistline indentations.$^{\ref{IMNS})}$\ess  When such nuclei collide with
their symmetry axes perpendicular to each other, the resulting configurations
are very compact, and little additional energy should be required to drive the
system inside its fission saddle point.  Reactions between such nuclei might
provide a path to the far side of the superheavy island.

\bigskip
\bigskip

\noindent{\bf 6.  Conclusion}

Three macroscopic-microscopic models have been used recently to calculate the
ground-state masses and deformations of nuclei throughout our known chart and
beyond.  One of these models can be extrapolated to new regions of nuclei whose
masses were not included in the original adjustment with considerable
confidence, and the other two can be extrapolated to new regions when some
caution is exercised.  Recent developments in these models are providing new
physical insight in several areas.  Macroscopic-microscopic models have also
correctly predicted the existence and location of a rock of deformed metastable
superheavy nuclei near $^{272}$110 that has recently been discovered.  Nuclear
ground-state masses and deformations will continue to provide an invaluable
testing ground for nuclear many-body theories.  The future challenge is for
\linebreak the fully microscopic theories to predict these quantities with
comparable or greater accuracy.

We are grateful to D.~G. Madland for stimulating discussions on relativistic
mean-field theory and chiral symmetry while writing this review.  This work was
supported by the U.~S. Department of Energy.

\fig me3d arles/me3dw 392 {Calculated microscopic
correction$^{\ref{MNMS},\ref{MNK})}$ in the vicinity of the recently discovered
rock of metastable superheavy nuclei near $^{272}$110.  An island near
$^{290}$110 beyond our present horizon is also predicted by these
calculations.}

\baselineskip=14.5pt
\bigskip
\bigskip
\noindent {\bf References}
\vspace{3.81pt}
\ben
\vspace{-6.5pt}

\item \label{QF} P.~Quentin and H.~Flocard, Ann.\ Rev.\ Nucl.\ Part.\ Sci.\
{\bf 28\/} (1978) 523.

\vspace{-6.5pt}

\item \label{SW} B.~D. Serot and J.~D. Walecka, Adv.\ Nucl.\ Phys.\ {\bf 16\/}
(1986) 1.
\vspace{-6.5pt}

\item \label{R} P.~G. Reinhard, Rep.\ Prog.\ Phys.\ {\bf 52\/} (1989) 439.
\vspace{-6.5pt}

\item \label{NHM} B.~A. Nikolaus, T.~Hoch, and D.~G. Madland, Phys.\ Rev.\ C
{\bf 46\/} (1992) 1757.
\vspace{-6.5pt}

\item \label{ST} Y. Sugahara and H. Toki, Nucl.\ Phys.\ A {\bf 579\/} (1994)
557.
\vspace{-6.5pt}

\item \label{LSHKR} G.~A. Lalazissis, M.~M. Sharma, W.~Hillebrandt, J.~K\"onig,
and P.~Ring, these \vspace{-6.5pt} Proceedings.

\item \label{W} S.~Weinberg, Phys.\ Lett.\ B {\bf 251\/} (1990) 288.
\vspace{-6.5pt}

\item \label{L} B.~W. Lynn, Nucl.\ Phys.\ B {\bf 402\/} (1993) 281.
\vspace{-6.5pt}

\item \label{MF} D.~G. Madland and J.~L. Friar, private communication (1995).
\vspace{-6.5pt}

\item \label{MNMS} P.~M\"oller, J.~R. Nix, W.~D. Myers, and W.~J. Swiatecki,
Atomic Data Nucl.\ Data Tables {\bf 59\/} (1995) 185.
\vspace{-6.5pt}

\item \label{MNK} P.~M\"oller, J.~R. Nix, and K.~L. Kratz, Atomic Data Nucl.\
Data Tables, to be published.
\vspace{-6.5pt}

\item \label{APDT} Y.~Aboussir, J.~M. Pearson, A.~K. Dutta, and F. Tondeur,
Nucl.\ Phys.\ A {\bf 549\/} (1992) \vspace{-6.5pt} 155.

\item \label{MS} W.~D. Myers and W.~J. Swiatecki, Nucl.\ Phys.\ A, to be
published.
\vspace{-6.5pt}

\item \label{MSb} W.~D. Myers and W.~J. Swiatecki, Lawrence Berkeley Laboratory
Report No.\ LBL-36803 (1994).
\vspace{-6.5pt}

\item \label{MSc} W.~D. Myers and W.~J. Swiatecki, ``The 1994 Thomas-Fermi
Model,'' these \vspace{-6.5pt} Proceedings.

\item \label{vGHT} H.~von~Groote, E.~R. Hilf, and K.~Takahashi, Atomic Data
Nucl.\ Data Tables {\bf 17\/} (1976) 418.
\vspace{-6.5pt}

\item \label{DZ} J.~Duflo and A.~P. Zuker, Phys.\ Rev.\ C, to be published.
\vspace{-6.5pt}

\item \label{GCPB} K.~A. Gernoth, J.~W. Clark, J.~S. Prater, and H.~Bohr,
Phys.\ Lett.\ B {\bf 300\/} (1993) 1.
\vspace{-6.5pt}

\item \label{S} V.~M. Strutinsky, Nucl.\ Phys.\ A {\bf 122\/} (1968) 1.
\vspace{-6.5pt}

\item \label{A} G.~Audi, Midstream Atomic Mass Evaluation, private
communication (1989), with four revisions.
\vspace{-6.5pt}

\item \label{AW} G.~Audi and A.~H. Wapstra, Nucl.\ Phys.\ A {\bf 565\/} (1993)
1.
\vspace{-6.5pt}

\item \label{SBNS} W.~Stocker, J.~Bartel, J.~R. Nix, and A.~J. Sierk, Nucl.\
Phys.\ A {\bf 489\/} (1988) 252.
\vspace{-6.5pt}

\item \label{MSd} W.~D. Myers and W.~J. Swiatecki, ``The Congruence Energy:\ a
Contribution to Nuclear Masses and Deformation Energies,'' these Proceedings.
\vspace{-6.5pt}

\item \label{MNMSb} P.~M\"oller, J.~R. Nix, W.~D. Myers, and W.~J. Swiatecki,
Nucl.\ Phys.\ A {\bf 536\/} (1992) 61.
\vspace{-6.5pt}

\item \label{MLN} P.~M\"oller, G.~A. Leander, and J.~R. Nix, Z. Phys.\ A {\bf
323\/} (1986) 41.
\vspace{-6.5pt}

\item \label{MN} P.~M\"oller and J.~R. Nix, Nucl.\ Phys.\ A {\bf 549\/} (1992)
84.  \vspace{-6.5pt}

\item \label{MNb} P.~M\"oller and J.~R. Nix, J. Phys.\ G: Nucl.\ Part.\ Phys.\
{\bf 20\/} (1994) 1681.
\vspace{-6.5pt}

\item \label{PS} Z.~Patyk and A.~Sobiczewski, Phys.\ Lett.\ B {\bf 256\/}
(1991) 307.
\vspace{-6.5pt}

\item \label{CS} S.~\'Cwiok and A.~Sobiczewski, Z. Phys.\ A {\bf 342\/} (1992)
203.
\vspace{-6.5pt}

\item \label{CHN} S.~\'Cwiok, S.~Hofmann, and W.~Nazarewicz, Nucl.\ Phys.\ A
{\bf 573\/} (1994) 356.
\vspace{-6.5pt}

\item \label{H+} S.~Hofmann et al., Z. Phys.\ A  {\bf 350\/} (1995) 277.
\vspace{-6.5pt}

\item \label{H+b} S.~Hofmann et al., Z. Phys.\ A  {\bf 350\/} (1995) 281.
\vspace{-6.5pt}

\item \label{H} S.~Hofmann, these Proceedings.
\vspace{-6.5pt}

\item \label{L+} Yu.~A. Lazarev et al., Phys.\ Rev.\ Lett.\ {\bf 73\/} (1994)
624.
\vspace{-6.5pt}

\item \label{O} Yu.~Ts.\ Oganessian, Nucl.\ Phys.\ A  {\bf 583\/} (1995) 823c.
\vspace{-6.5pt}

\item \label{Ob} Yu.~Ts.\ Oganessian, these Proceedings.
\vspace{-6.5pt}

\item \label{IMNS} A.~Iwamoto, P.~M\"oller, J.~R. Nix, and H.~Sagawa, Nucl.\
Phys.\ A, to be published.
\een

\end{document}